\pdfoutput=1 
\RequirePackage{amsmath}
\documentclass[envcountsame,runningheads]{llncs}
\usepackage{cite} 

\newif\iflong   
\newif\ifmorecites
\newif\ifanonymous
\newif\ifcomments   

 \commentstrue

\providecommand{\keywords}[1]{\textbf{Keywords:} #1}

\usepackage[T1]{fontenc}    
\usepackage[english]{babel} 
\usepackage{csquotes}       
\usepackage{amsmath}        
\usepackage{dsfont}         
\usepackage{upgreek}        
\usepackage{mathtools}      
\usepackage{enumitem}       
\usepackage{graphicx}       
\graphicspath{ {./images/} }
\usepackage{booktabs}       
\usepackage{array,multirow} 
\usepackage{tabularx}       
\usepackage{amsfonts}       
\usepackage{hyperref} 
\usepackage[dvipsnames]{xcolor}
\definecolor{lightgray204}{RGB}{204,204,204}

\hypersetup{                
    colorlinks,
    linkcolor={red!50!black},
    citecolor={blue!50!black},
    urlcolor={blue!80!black}
}
\usepackage{etoolbox}\appto\UrlBreaks{\do\-}
\usepackage[capitalise]{cleveref}       

\usepackage{xparse}
 \usepackage{pgfplots}

\ifcomments\fi  
\usepackage[colorinlistoftodos,prependcaption,textsize=tiny,textwidth=\marginparwidth]{todonotes}

\newcommand{\yotam}[1]{\ifcomments {\todo[color=green!40,inline]{Yotam: #1}} \fi}

\usepackage{thmtools}
\usepackage{thm-restate}   

\spnewtheorem{claim}{Claim}{\bfseries}{\itshape}

\spnewtheorem{example}{Example}{\bfseries}{\itshape}
\spnewtheorem{fact}{Fact}[definition]{\bfseries}{\itshape}
\spnewtheorem{conclusion}{Conclusion}{\bfseries}{\itshape}
\spnewtheorem{prop}{Proposition}{\bfseries}{\itshape}
\usepackage{chngcntr}
\counterwithin{definition}{section}
\counterwithin{theorem}{section}
\counterwithin{prop}{section}
\counterwithin{claim}{section}
\counterwithin{fact}{section}
\counterwithin{lemma}{section}
\counterwithin{corollary}{section}
\counterwithin{example}{section}
\counterwithin{remark}{section}

\usepackage{etoolbox} 
\AtEndEnvironment{proof}{\phantom{}\qed} 

\crefname{definition}{Def.}{Defs.}
\crefname{theorem}{Theorem}{Theorems}
\crefname{claim}{Claim}{Claims}
\crefname{fact}{Fact}{Fact}
\crefname{lemma}{Lemma}{Lemmas}
\crefname{prop}{Proposition}{Propositions}

\crefname{corollary}{Corollary}{Corollaries}
\crefname{example}{Example}{Examples}
\crefname{remark}{Remark}{Remarks}
\crefname{code}{Code}{Code}

\newcommand{\define}{\stackrel{\mathclap{\mbox{def}}}{=}}

\usepackage[symbols,acronym,nonumberlist,nogroupskip,section=subsection,numberedsection,stylemods={mcols,longbooktabs}]{glossaries-extra}
\makenoidxglossaries

\glssetcategoryattribute{acronym}{nohyper}{true} 
\setabbreviationstyle[acronym]{long-short}  
\newacronym{EVM}{EVM}{Ethereum virtual machine}
\newacronym{DeFi}{DeFi}{decentralized finance}
\newacronym{PoW}{PoW}{Proof-of-Work}
\newacronym{PoS}{PoS}{Proof-of-Stake}
\newacronym{MEV}{MEV}{miner-extractable value}
\newacronym{block-DAG}{block-DAGs}{block directed-acyclic-graph}
\newacronym{DAA}{DAA}{difficulty-adjustment algorithm}
\newacronym{MDP}{MDP}{Markov decision-process}
\newacronym{DQL}{DQL}{Deep-Q-learning}
\newacronym{RL}{RL}{reinforcement learning}
\newacronym{ML}{ML}{machine learning}
\newacronym{AI}{AIs}{artificial intelligence}
\newacronym{PDF}{PDF}{probability density function}
\newacronym{CDF}{CDF}{cumulative density function}
\newacronym{AMM}{AMM}{automated market maker}
\newacronym{USD}{USD}{United States Dollar}
\newacronym{IP}{IP}{Internet Protocol}
\newacronym{LP}{LP}{Liquidity Provider}
\newacronym{LT}{LT}{Liquidity Taker}
\newacronym{APY}{APY}{annual percentage yield}
\newacronym{PID}{PID}{proportional integral derivative}
\newacronym{UTXO}{UTXO}{unspent transaction output}
\newacronym{YAML}{YAML}{YAML Ain't Markup Language}
\newacronym{TD}{TD}{total difficulty}
\newacronym{geth}{geth}{Go Ethereum}
\newacronym{WETH}{WETH}{Wrapped Ethereum}
\newacronym{ASIC}{ASIC}{Application Specific Integrated Circuit}
\newacronym{RPC}{RPC}{remote procedure call}
\newacronym{RUM}{RUM}{riskless uncle maker}
\newacronym{PUM}{PUM}{preemptive uncle maker}
\newacronym{URL}{URL}{uniform resource locator}
\newacronym{SSD}{SSD}{solid state drive}
\newacronym{EIP}{EIP}{Ethereum improvement proposal}
\newacronym{CPU}{CPU}{central processing unit}
\newacronym{RAM}{RAM}{random-access memory}
\newacronym{mempool}{mempool}{memory pool}
\newacronym{TFM}{TFM}{transaction fee mechanism}
\newacronym{iid}{i.i.d.}{independent and identically distributed}
\newacronym{wrt}{w.r.t.}{with regard to}
\newacronym{wlog}{w.l.o.g.}{without loss of generality}
\newacronym{TTL}{TTL}{time to live}
\newacronym{DSIC}{DSIC}{dominant strategy incentive-compatible}
\newacronym{EPIC}{EPIC}{ex-post incentive-compatible}
\newacronym{BIC}{BIC}{Bayesian incentive-compatible}
\newacronym{MMIC}{MMIC}{myopic miner incentive-compatible}
\newacronym{FMIC}{FMIC}{farsighted miner incentive-compatible}
\newacronym{OCA}{OCA}{off-chain-agreement}
\newacronym{SCP}{SCP}{side-contract-proof}
\newacronym{PABGA}{PABGA}{pay-as-bid greedy auction}
\newacronym{SPA}{SPA}{second price auction}
\newacronym{UPGA}{UPGA}{uniform-price greedy auction}
\newacronym{BNE}{BNE}{Bayesian-Nash equilibrium}
\newacronym{EPIR}{EPIR}{ex-post individually rational}
\newacronym{EPBB}{EPBB}{ex-post burn balanced}
\newacronym{GTA}{GTA}{good toy auction}
\newacronym{GTFBA}{GTFBA}{good toy finite-blocksize auction}
\newacronym{QoS}{QoS}{quality of service}

\glsxtrnewsymbol[description={
    An auction.
}]{auction}{
    \ensuremath{A}
}
\newcommand{\auction}{{\gls[hyper=false]{auction}}}

\glsxtrnewsymbol[description={
    A transaction.
}]{tx}{
    \ensuremath{tx}
}
\newcommand{\tx}{{\gls[hyper=false]{tx}}}

\glsxtrnewsymbol[description={
    Payment rule.
}]{pay}{
    \ensuremath{p}
}
\newcommand{\pay}{{\gls[hyper=false]{pay}}}

\glsxtrnewsymbol[description={
    Burning rule.
}]{burn}{
    \ensuremath{\beta}
}
\newcommand{\burn}{{\gls[hyper=false]{burn}}}

\glsxtrnewsymbol[description={
    Update rule.
}]{update}{
    \ensuremath{\upsilon}
}

\glsxtrnewsymbol[description={
    \gls[hyper=false]{TTL} of a transaction.
}]{ttl}{
    \ensuremath{t}
}

\glsxtrnewsymbol[description={
    Reserve price of an auction.
}]{reserve}{
    \ensuremath{r}
}
\newcommand{\reserve}{{\gls[hyper=false]{reserve}}}

\glsxtrnewsymbol[description={
    \gls[hyper=false]{TTL} of past transactions.  
}]{mempoolSymb}{
    \ensuremath{\mu}
}

\glsxtrnewsymbol[description={
    Immediacy ratio for our non-myopic allocation rule.
}]{imratio}{
    \ensuremath{\ell}
}

\glsxtrnewsymbol[description={
    Transaction schedule function.
}]{adversary}{
    \ensuremath{\psi}
}

\glsxtrnewsymbol[description={
    Allocation function.
}]{allocation}{
    \ensuremath{x}
}
\newcommand{\alloc}{{\gls[hyper=false]{allocation}}}

\glsxtrnewsymbol[description={
    Transaction fee of some transaction, in tokens.
}]{fee}{
    \ensuremath{b}
}
\newcommand{\fee}{{\gls[hyper=false]{fee}}}

\glsxtrnewsymbol[description={
    Maximal \gls[hyper=false]{TTL} of transactions in the blockchain, in blocks.
}]{maxttl}{
    \ensuremath{d}
}

\glsxtrnewsymbol[description={
    Predefined maximal block-size, in bytes.
}]{blocksize}{
    \ensuremath{\mathcal{B}}
}
\newcommand{\blocksize}{{\gls[hyper=false]{blocksize}}}

\glsxtrnewsymbol[description={
    Miner's relative mining power, as a fraction.
}]{minerRatio}{
    \ensuremath{\alpha}
}

\glsxtrnewsymbol[description={
    Miner discount factor.
}]{discount}{
    \ensuremath{\lambda}
}

\glsxtrnewsymbol[description={
    Miner planning horizon.
}]{horizon}{
    \ensuremath{T}
}

\glsxtrnewsymbol[description={
    Miner revenue.
}]{revenue}{
    \ensuremath{u}
}
\newcommand{\revenue}{{\gls[hyper=false]{revenue}}}

\glsxtrnewsymbol[description={
    Number of all bids with $b_i \geq \reserve$.
}]{bidsMoreReserve}{
    \ensuremath{S_{\geq \reserve}}
}

\newcommand{\feeDomain}{\ensuremath{\Phi}}
\newcommand{\txDomain}{\ensuremath{\feeDomain}}

\NewDocumentCommand{\expect}{ e{_} s o >{\SplitArgument{1}{|}}m }{%
  \operatorname{E}
  \IfValueT{#1}{{\!}_{#1}}
  \IfBooleanTF{#2}{
    \expectarg*{\expectvar#4}%
  }{
    \IfNoValueTF{#3}{
      \expectarg{\expectvar#4}%
    }{
      \expectarg[#3]{\expectvar#4}%
    }%
  }%
}
\NewDocumentCommand{\expectvar}{mm}{%
  #1\IfValueT{#2}{\nonscript\;\delimsize\vert\nonscript\;#2}%
}
\DeclarePairedDelimiterX{\expectarg}[1]{[}{]}{#1}

\usepackage{csquotes}
\usepackage{pgfplots}

\pgfdeclarelayer{background}
\pgfsetlayers{background,main}

\title{Incentive-compatible Collusion-resistance via Posted Prices}
\date{}
\ifanonymous\author{}\institute{}\else

\author{Mathues V.X. Ferreira\inst{1} \and Yotam Gafni\inst{2} 
 \and Max Resnick\inst{3}}

\institute{
University of Virginia, \\
\email{matheus@virginia.edu} \and
Weizmann Institute,\\
\email{yotam.gafni@gmail.com}
\and
Special Mechanisms Group \email{max@specialmechanisms.org}}

\fi

\pgfplotsset{compat=1.18}

\begin{document}

\maketitle
\begin{abstract}
 We consider a refinement to the notions of collusion-resistance in transaction fee mechanisms. In particular, we require that the collusion is by itself incentive-compatible and individually rational to all of its participants. We then study the structural properties of these notions, and importantly, characterize the class of collusion-resistant and incentive-compatible transaction fee mechanisms in the single bidder case, and show that this is exactly the class of posted-price where the price is not too prohibitive. We analyze welfare and revenue implications, as well as the shape of the solution space, for both regular and non-regular distributions. 
\end{abstract}

\section{Introduction}

Collusion-resistance is a long studied problem in the economic literature \cite{Stigler1964}. Recently, it has also gained importance as a primitive in Blockchain mechanism design \cite{roughgarden2021transaction,fox2023censorship}, in particular the design of transaction fee mechanisms. 
Roughgarden \cite{roughgarden2021transaction} defines the notion of an off-chain agreement (OCA), and shows that being OCA-proof is equivalent to maximizing the joint utility of the winning coalition together with the miner. The work then presents a main axiomatic question, of whether it is possible to arrive at a mechanism which is incentive-compatible for users (DSIC), incentive-compatible for miners (MMIC), and OCA-proof. This has recently been answered in the negative \cite{gafni2024barriers,chung2024collusionresilience}. 
Chung \& Shi \cite{chung2022foundations} discuss a collusion notion they call $c$-SCP (side-contract proof), which means that no coalition of $c$ users and the miner can benefit from deviation. They show that DSIC+$1$-SCP is only achieved by the trivial mechanism, and in particular the revenue to the miner must be $0$. Gafni \& Yaish \cite{gafni2024barriers} show that OCA and SCP are different notions, where SCP is strictly stronger. This paints a picture where it is impossible to achieve collusion resistance in transaction fee mechanisms, and is exacerbated by further results under relaxing assumptions \cite{shi2023cryptography,chen2023bayesian}. 

Our main conceptual observation is that the collusion notions (OCA, SCP) are too strict in their current form. Most importantly, they assume complete trust between the colluders, in the sense that the colluders do not deviate or misuse the established collusion itself. This can may be best demonstrated by the example of avoiding a reserve price. 

\begin{example}[A motivating example I, informal]
\label{ex:motivating1}

Consider a seller that sells an item to a single buyer. Let us assume that it is defined by the protocol that the item needs to be sold for $10$ USD. However, the buyer only values it at $5$ USD. Then, the seller and buyer could collude against the protocol in the following fashion: The buyer would declare that its value for the item is $10$ USD and buy the item. Then, the seller would ``cashback'' the buyer for $6$ USD. The seller ends up with a revenue of $4$, and the buyer with a utility of $1$, instead of $0$ in the original situation where the protocol does not let the purchase go through.

\textit{However}, now consider a user with a value of $20$ USD. They know the seller's proclivity to cashbacks, and so decide to declare that they only have a value of $4$ USD for the item. Since the miner has no way to differentiate this buyer from the buyer who truly only has a value of $5$, they go forward with the same cashback scheme, i.e., selling the item for effectively $4$ USD. The seller has a revenue of $4$, the buyer a utility of $16$, whereas in the original situation both have a utility of $10$. Thus, the collusion was \textit{exploitable} by users of some type, and we say it is not \textit{incentive-compatible}.
\end{example}

\begin{example}[A motivating example II, informal]
\label{ex:motivating2}

Consider a seller that sells an item through the ``tipless mechanism'' \cite{roughgarden2021transaction}, i.e., for some constant price and burn. A disadvantage of the tipless mechanism is that the item is not guaranteed to go to the bidder that values it the most. To mitigate that, the seller decides to hold a second-price auction among all the bidders with bid higher than the posted price. 

\textit{However}, the miner can exploit the collusion by submitting additional fake bids, thus raising the payment of the winning colluder. 
We thus say this suggested collusion is not \textit{incentive-compatible} (in this case, for the miner).
\end{example}

Both examples are captured by our Definitions~\ref{def:ic_collusion},\ref{def:ir_collusion}. By imposing the incentive-compatibility restrictions, we require that the collusion is somehow well-designed, or complete, and does not leave out ``holes'' that can be exploited by the parties to the collusion. However, even an incentive-compatible collusion can be infeasible, as we demonstrate in our final example. 

\begin{example}[A motivating example I, revisited]
\label{ex:motivating3}

Consider a seller that sells an item to a single buyer. Let us assume that it is defined by a protocol that the item needs to be sold for $10$ USD. 
We want to define an incentive-compatible collusion that allows a buyer with a value of $5$ USD for the item to buy it. The format we suggest is the following: Any bidder bids her true value when setting up the collusion mechanism. If the value is $5$ USD or more, the bidder bids $10$ USD in the actual TFM. The item is then allocated to her (as per the protocol), and the miner cashes her back $5$ USD. Unlike Example~\ref{ex:motivating1}, this collusion mechanism is incentive-compatible because bidders such as the $20$ USD value bidder do not need to ``fake'' their way into the collusion, but they are already incorporated into it. Moreover, we can observe that the collusion in fact maps an incentive-compatible TFM (posted price at $10$ USD) into another incentive-compatible TFM (posted price at $5$ USD). 

\textit{However}, it is still not necessarily a feasible collusion. For the user, it is definitely better to buy the item for $5$ USD instead of $10$ USD. For the miner, it can sometimes be better, if $5$ USD is a better price that allows for higher demand, enough so to compensate for the lower price, but it can also sometimes be worse. This can be decided given a Bayesian distribution over the possible bidders. We thus require that in expectation, each party to the collusion is better off by it. 
\end{example}



A similar observation was recently made by \cite{weinbergRevisiting}:

\begin{displayquote}
```We observe that prior definitions (OCA proofness, $1$-SCP, and our strong collusion proofness) mostly seem to capture collusion \textit{between trusted entities} [...]'''
\end{displayquote}

\cite{gafni2024barriers} emphasize that in the traditional economic literature, there are many obstacles to implementing a successful collusion, including strategic deviation by the colluders, agreeing on profit-sharing in the coalitional formation game (see, e.g., \cite{coreCoalitionFormation}), and the limits of quick communication. We focus on the obstacle of trust between the players, given that there is a fixed take-it-or-leave-it proposed collusion. This is reasonable given the central role the miner plays in deciding allocations, at least with Proof-of-Stake as implemented in Ethereum (as the block proposer for the next block is decided in the beginning of the block production epoch). Generally, the full information / trustful collusion setting best fits a situation where the miner is an active participant in the market, and issues their own transactions while fulfilling a second role as validator. The trustless collusion setting best fits a situation where the miner is dedicated to the role of validator, but engages in collusion with bidders against the protocol to enhance their profits.

\subsection{Our Results}

First, we define the incentive-compatible (IC) and individually rational (IR) version of collusion. We then fully characterize the single bidder case, following the structure of the characterization in the paper by Gafni \& Yaish \cite{gafni2024barriers}. Even in this limited single bidder case, we can already see the difference once IC+IR are introduced into the collusion notions. While Lemma~3.5 of \cite{gafni2024barriers} shows that any single bidder DSIC+MMIC+OCA mechanism must have $0$ miner revenue, we show that DSIC+MMIC+(IC+IR)-OCA mechanisms offer all the range of Pareto-optimal auctions between $0$ miner revenue and optimal miner revenue, whether the bidder valuation is drawn from a regular distribution (Theorem~\ref{thm:single_bidder_regular}), or from a non-regular distribution (Theorem~\ref{thm:single_bidder_nonregular}). However, the regular case is ``cleaner'' as it simply yields a contiguous interval of feasible posted prices, while the non-regular may have discontinuities (Example~\ref{ex:nonregular_discontinuity}).

\section{Model and Preliminaries}
\label{sec:Model}
We follow the standard model used by the literature for \gls{TFM}s \cite{lavi2019redesigning},\cite{roughgarden2021transaction},\cite{roughgarden2020transaction},\cite{chung2023foundations},\cite{shi2023what}. The main technical assumption we make, which we use to simplify the discussion, is that the auction is a \textit{single-item} auction, i.e., that the block size is $1$. We interchangeably use the terms \textit{auction}, \textit{mechanism}, or \gls{TFM} to refer to the object of our discussion, depending on what is most appropriate in the context. 

\subsection{Transaction Fee Mechanisms}
A \gls{TFM} $\auction$ consists of an \emph{allocation rule} $\alloc^\auction$, a \emph{payment rule} $\pay^\auction$ and a \emph{burning rule} $\burn^\auction$.
For brevity, when it is clear from context, we omit the $\auction$ identifier from the notations.

\subsubsection*{Allocation rule}
An allocation rule defines the mechanism's allocation of transactions to the upcoming block, as intended by the mechanism's designers.
\begin{definition}[Allocation rule]
    \label{def:AllocFunc}
    A \textit{deterministic} allocation rule $\alloc_n^\auction:R_+^n \rightarrow \{0,1\}^n$ of auction $\auction$ defines which transaction should be included in the upcoming block by $\auction$. I.e., we require that $\forall \mathbf{b} \in R_+^n: \sum_i [\alloc_n^\auction(\mathbf{b})]_i \leq 1$.
    %
\end{definition}
\begin{remark}
    In our analysis, we omit the identifier $n$ and write $\alloc^\auction$.
    We assume that the correct $\alloc_n^\auction$ is used, based on the length $n$ of the vector of bids $\mathbf{b}$. We also use the following auxiliary notation: 
$\tilde{\alloc}^\auction:R_+^n\rightarrow [n] \cup \{\emptyset\}$ 
    to specify the unique bidder who receives the item, or $\emptyset$ if the item remains un-allocated.
    I.e., if $\tilde{\alloc}^\auction(\mathbf{b}) = i$, then:
    \begin{equation*}
        \alloc^\auction(\mathbf{b}) = \{0, \ldots, 0, \underbrace{1}_{i\text{-th index}}, 0, \ldots, 0\}.
    \end{equation*}

\end{remark}

\subsubsection*{Payment \& burn}
The payment and burn rules respectively define the amount of fees paid by each transaction when included in a block, and how much of each payment is ``burnt'' and taken out of circulation instead of being given to the miner, with both rules enforced by the mechanism.
\begin{definition}[Payment rule]
    The payment rule $\mathbf{\pay^\auction}:R_+^n\rightarrow R_+^n$ in auction $\auction$ receives bids $b_1, \ldots, b_n$, and outputs a payment vector with a payment for each bidder $i$.
\end{definition}
\begin{definition}[Burning rule]
    The burning rule $\mathbf{\burn^{\auction}}:R_+^n\rightarrow R_+^n$ in auction $\auction$ receives bids $b_1, \ldots, b_n$, and and outputs a burn vector with a burn for each bidder $i$, meaning the amount of funds out of $b_i$'s payment that is taken out of circulation.
\end{definition}

\begin{remark}
For $\zeta \in \{\pay^\auction, \burn^\auction\}$, we use $\tilde{\zeta}:R_+^n \rightarrow R$ to denote the payment or burn of the \textit{winner}, i.e., if $\tilde{\alloc}^\auction(\mathbf{b}) = i$, then: $\zeta(\mathbf{b}) = \{0, \ldots, 0, \underbrace{\tilde{\zeta}(\mathbf{b})}_{i\text{-th index}}, 0, \ldots, 0\}$.
\end{remark}


\subsubsection*{Basic definitions}
We proceed with several definitions that are used throughout our analysis.

\begin{definition}[Agent Utilities]
    \label{def:utilities}
    Let auction $\auction = (\alloc, \pay, \burn)$. Consider $n$ bidders with true valuations $\mathbf{v} \in R_+^n$ and $n'$ bids 
    $\mathbf{b} \in R^{n'}$. 
    We assume that $\mathbf{b}$ is indexed the same as $\mathbf{v}$, meaning that the first $n$ bids in $\mathbf{b}$ correspond to the bids in $\mathbf{v}$ (possibly as $0$ if omitted), and so w.l.o.g. we consider $n' \geq n$. We consider the $n' - n$ bidders that are in $\mathbf{b}$ but not in $\mathbf{v}$ as the miner's ``fake bidders'', and denote them by $\mathbf{b}_F$. 
    Then, the utilities of the various agents are defined as follows.
    \begin{itemize}
    \item  \textit{Bidder Utility:}
    \begin{equation}
    \label{eq:bidder_util}
        u_i(\fee_i, \mathbf{\fee_{-i}} ; v_i)
        \define
            v_i \cdot \alloc(b_i, \mathbf{b}_{-i}) - \pay\left(\fee_i,\mathbf{\fee_{-i}}\right).
      \end{equation}

    \item \textit{Miner Utility:}
    \begin{equation}
    \label{eq:miner_util}
    \begin{split}
    u_{miner} (\mathbf{\fee} ; \mathbf{v})
    &
    \define
    \sum_{i=1}^{n'} \left(\pay(b_i, \mathbf{b}_{-i}) - \burn(b_i, \mathbf{b}_{-i}) \right) - \sum_{i=n+1}^{n'} u_i(\fee_i, \mathbf{\fee_{-i}} ; 0)
    \\&
    =
    \sum_{i=1}^n \left(\pay(b_i, \mathbf{b}_{-i}) - \burn(b_i, \mathbf{b}_{-i}) \right) - \sum_{i=n+1}^{n'} \burn(b_i, \mathbf{b}_{-i}).
    \end{split}
    \end{equation}

    \item
    \textit{Joint Utility:}
    \begin{equation}
    \label{eq:joint_util}
    \begin{split}
    u_{joint}(\mathbf{b} ; \mathbf{v})
    &
    \define
    u_{miner}(\mathbf{b} ; \mathbf{v}) + \sum_{i=1}^n u_i(b_i, \mathbf{b}_{-i} ; v_i)
    \\&
    =
    \sum_{i=1}^n \left( v_i \alloc(b_i, \mathbf{b}_{-i}) - \burn(b_i, \mathbf{b}_{-i})\right) - \sum_{i=n+1}^{n'} \burn(b_i, \mathbf{b}_{-i}).
    \end{split}
    \end{equation}
    \end{itemize}
\end{definition}

\begin{definition}[Basic Auction Properties]
We say an auction is:
\begin{itemize}
    \item \textit{Individually Rational}, if the expected payment by an agent does not exceed its bid if it is allocated, and $0$ if it is not.
    Formally, for any $\mathbf{b} \in R_+^n$ and for each bidder $i$, we have
    \begin{equation}
    \label{eq:IR_util}
    u_i(b_i, \mathbf{b}_{-i} ; b_i) \geq 0.
    \end{equation}
    If we develop this condition given the explicit form of bidder utility in \cref{eq:bidder_util}, we get:
    \begin{equation}
    \label{eq:IR}
    \alloc(b_i, \mathbf{b}_{-i}) \cdot b_i \geq \pay(b_i, \mathbf{b}_{-i}).
    \end{equation}

    \item \textit{Burn-Balanced}, if the expected burn is non-negative and does not exceed bidder payments.
    Formally, for any $\mathbf{b} \in R_+^n$ and for each bidder $i$, we have
    \begin{equation}
    \label{eq:BB}
    \pay(b_i, \mathbf{b}_{-i}) \geq \burn(b_i, \mathbf{b}_{-i}) \geq 0,
    \end{equation}

    \item \textit{Anonymous}, if the allocation, payment, and burn rules are agnostic to permutations over the bidders.
    Formally, for any vector of bids $\mathbf{b} \in R_+^n$, permutation $\pi \in S_n$, and rule $x \in \{\alloc, \pay, \burn\}$, we have
    $x(\pi(\mathbf{b})) = \pi(x(\mathbf{b}))$.
    For anonymous auctions, we treat bids as an unordered set.
\end{itemize}
\end{definition}

\subsection{The TFM Desiderata}
\label{sec:Desiderata}

\begin{definition}[\Glsxtrfull{DSIC}]
    \label{def:DSIC}
    An auction is \gls{DSIC} if it is always weakly better for bidders to declare their true values.
    Thus, for any bidder $i$ with true value $v_i$, for any bid $b_i$, and for any bids $\mathbf{b_{-i}}$ by all other bidders, we have
    $
    u_i(v_i, \mathbf{b_{-i}} ; v_i) \geq u_i(b_i, \mathbf{b_{-i}} ; v_i),
    $
    and there is such $\mathbf{b_{-i}}$ so that the inequality is strict. 
\end{definition}

\begin{definition}[Miner Strategy Function, \Glsxtrfull{MMIC}]
\label{def:miner_strategy}

    A \textit{miner strategy function} $\mu:R^n\rightarrow R^{n'}$ is defined such that for any $\mathbf{v}\in R^n$ (i.e., a vector of bids of length $n$), it outputs a vector $\mathbf{b}\in R^{n'}$ of some length $n'\geq n$, so that for any $1\leq i \leq n$, $b_i \in \{0, v_i\}$.
    I.e., it has the choice whether to include or omit any of the original $n$ bids, as well as add $n' - n$ fake bids. Importantly, a miner cannot \textit{change} the real bidders' bids. 

    A \gls{TFM} is \gls{MMIC} if a miner cannot strictly increase its revenue by deviating from the intended allocation rule, or by introducing ``fake'' miner-created bids.
    Formally, for any miner strategy $\mu$, and any valuation $\mathbf{v}$,
    \begin{equation}
    \label{eq:mmic_cond}
    u_{miner}(\mathbf{v} ; \mathbf{v}) \geq u_{miner}(\mu(\mathbf{v}) ; \mathbf{v}).
    \end{equation}
    
\end{definition}


\subsection{Incentive-Compatible and Individually-Rational Collusion}

\begin{definition}
A $k$-collusion is defined by a bids-changing function $c:R^n_+\rightarrow R^{n'}_+$, and a transfer function $t:R^n_+\rightarrow R^{n+1}$. We emphasize a few properties:
\begin{itemize}

\item $c$ preserves the coordinates of all the $n-k$ bidders not in the collusion through the transformation, 

\item For all the $n-k$ bidders not in the collusion, their transfer value $t_i$ is always $0$.

\item For any $\mathbf{b}$, $\sum_{i=1}^{n+1} t_i(\mathbf{b}) = 0$ (i.e., the transfers sum up to $0$ in each instance). Notice we include the miner in the last coordinate of the transfer function, and allow for omission / addition of fake bids by the miner in the bids-changing function (hence its domain is $R^{n'}_+$). 

\item The transfers satisfy ex-post individual rationality for the colluding bidders. Thus, for any colluding bidder $i$, and any bid vector $\mathbf{b}$, $u_i(c(\mathbf{b}) ; \mathbf{b}) + t_i(\mathbf{b}) \geq 0$.  

\end{itemize}



    







\end{definition}


\begin{definition}
\label{def:ic_collusion}
An incentive compatible (IC) collusion satisfies for each bidder $i$ in the collusion (either through altering their bid or through non-zero transfers), for any $\mathbf{b}$ so that $c$ is not the identity function or $t$ is not the $0$ function:

$u_i(c(\mathbf{b}) ; b_i) + t_i(\mathbf{b}) \geq u_i(c(b'_i, b_{-i}) ; b_i) + t_i(b'_i, b_{-i})$,

and for the miner, for any miner strategy function $\mu$ (as defined in Definition~\ref{def:miner_strategy}):

$u_{miner}(c(\mathbf{b}) ; \mathbf{b}) + t_{miner}(\mathbf{b}) \geq u_{miner}(c(\mu(\mathbf{b})) ; \mathbf{b}) + t_{miner}(\mu(\mathbf{b})).$

\end{definition}

\begin{definition}
\label{def:ir_collusion}
An individually rational (IR) $k$-collusion satisfies for each bidder $i$ in the collusion:

$E_{\mathbf{b}\sim B}[u_i(c(\mathbf{b}) ; b_i) + t_i(\mathbf{b})] \geq E_{\mathbf{b}\sim B}[u_i(\mathbf{b} ; b_i)]$,

and for the miner 

$E_{\mathbf{b}\sim B}[u_{miner}(c(\mathbf{b}) ; \mathbf{b}) + t_{n+1}(\mathbf{b})] \geq E_{\mathbf{b}\sim B}[u_{miner}(\mathbf{b} ; \mathbf{b})]$.

\end{definition}

\begin{remark}
Notice that we define the collusion individual rationality notion as ex-ante, rather than ex-interim (i.e., where it must be individually rational for the bidders to follow the collusion after knowing their value, whatever it may be). It is important to notice that since individual rationality is used to filter ``unreasonable'' collusions, choosing the lighter ex-ante notion is more restrictive, and so our positive results naturally hold in the ex-interim case as well.
\end{remark}

We are now ready to define the incentive-compatible and individually rational versions of the two prominent collusion notions in the transaction fee mechanism literature.

\begin{definition}
\textit{(IC+IR)-OCA-proof} mechanism is such that is resistant to any $n$-collusion, i.e., there is no IC+IR $n$-collusion against it.

\textit{(IC+IR)-SCP} mechanism is such that is resistant to any $k$-collusion, with $1 \leq k \leq n$.

\end{definition}

\section{The Single Bidder Case}

In the single bidder case, there is no difference between the collusion notions. 

\begin{example}[Motivating examples I-III, formal]

We revisit the examples we give in the introduction. 

In \textit{Example I}, we have $k = n = 1$ (i.e., a miner and a user colluding in a single-bidder auction). The auction is $x(b_1) = (b_1 \geq 10), \tilde{p}(b_1) = 10, \tilde{\beta}(b_1) = 0$. 

The collusion is:
$$c(\mathbf{b}) = \begin{cases} (10) & \mathbf{b} = (5) \\ \mathbf{b} & Otherwise\end{cases}, t(\mathbf{b}) = \begin{cases} (6,-6) & \mathbf{b} = (5) \\ (0,0) & Otherwise\end{cases}.$$

This collusion violates the IC condition for bidder $1$, as:
$$u_1(c(20) ; 20) + t_1(20) = u_1(20 ; 20) + 0 = 10 < u_1(c(5) ; 20) + t_1(5) = u_1(10 ; 20) + 6 = 16. $$

In \textit{Example II}, we have for any $k \leq n$, posted price $P$ and burn $B \leq P$, the auction:

$\tilde{x}(\mathbf{b}) = \min_i (b_i \geq P), \tilde{p}(\mathbf{b}) = P, \tilde{\beta}(\mathbf{b}) = B$. 

The collusion is:
\[
\begin{split}
& c(\mathbf{b}) = \begin{cases}\{b_1, 0, \ldots, 0\} & b_1 \geq P \\ \mathbf{b} & Otherwise,
\end{cases}, \\
& t(\mathbf{b}) = \begin{cases} (\min\{-b_2 + P,0\},\max\{b_2 - P, 0\}) & b_1 \geq P \\ (0,0) & Otherwise\end{cases},
\end{split}
\]

assuming w.l.o.g. that $b_1 > b_2 > \ldots b_k$. I.e., all colluding bidders besides the highest colluding bidder drop their bids, and the highest colluding bidder, if she wins, pays the miner the difference between the protocol's set price and the second highest colluding bidder. 

This collusion satisfies the IC condition for bidders, since for the colluding bidders it simulates (together with the auction $(x,p,\beta)$) the second-price auction with reserve $P$. It violates the IC condition for the miner since say we fix $k = 2, n = 3$, then in the case that only one of the colluding bidders $b_1$ shows up, the miner can \textit{pretend} in that case (as long as $b_1 > P$) that there is another colluding bidder $b_2$ with $b_1 > b_2 = \frac{P + b_1}{2} > P$, i.e., $\mu(b_1, b_3) = \begin{cases} (b_1, \frac{P + b_1}{2}, b_3) & b_1 > P \\ (b_1, b_3) & Otherwise,\end{cases}$. Then, we have 

$$u_{miner}(c(\mathbf{b}) ; \mathbf{b}) + t_{miner}(\mathbf{b}) = P < \frac{P+b_1}{2} = u_{miner}(c(\mu(\mathbf{b})) ; \mathbf{b}) + t_{miner}(\mu(\mathbf{b})). $$

In \textit{Example III}, we again have $k = n = 1$, and the auction $x(b_1) = (b_1 \geq 10), \tilde{p}(b_1) = 10, \tilde{\beta}(b_1) = 0$. 

The collusion in this example is:
$$c(\mathbf{b}) = \begin{cases} (10) & b_1 \mathbf{\geq} 5, \\\mathbf{b} & Otherwise\end{cases}, t(\mathbf{b}) = \begin{cases} (5,-5) & b_1 \geq 5 \\ (0,0) & Otherwise\end{cases}.$$

Then, 

$$u_1(c(b_1) ; v_1) + t_1(b_1) = \begin{cases} u_1(10 ; v_1) + 5 & b_1 \geq 5\\ 
u_1(b_1 ; v_1) & b_1 < 5\end{cases} = \begin{cases} v_1 - 10 + 5 & b_1 \geq 5\\ 
0 & b_1 < 5\end{cases},$$
and indeed it is optimal to declare the true $v_1$. It is also incentive-compatible for the miner not to omit the bid $b_1$ or add its own bids because it has utility $0$ in case $c(b_1)$ is not allocated, and a non-negative utility if $c(b_1)$ is allocated. 

\end{example}

\begin{lemma}
An IR collusion must weakly increase the expected social welfare of the $k$ colluders and the miner.
\end{lemma}
\begin{proof}
Simply sum over the IR conditions:

\[
\begin{split}
& E_{\mathbf{b}\sim B}[SW_{colluders}(c(\mathbf{b}))] = \sum_i E_{\mathbf{b}\sim B}[a_i(c(\mathbf{b})) \cdot v_i] = \\
& \sum_i E_{\mathbf{b}\sim B}[u_i(c(\mathbf{b})) + t_i(\mathbf{b})] + E_{\mathbf{b}\sim B}[p(c(\mathbf{b})) + t_{n+1}(\mathbf{b})] \geq \\
& E_{\mathbf{b}\sim B}[u_i(\mathbf{b})] + E_{\mathbf{b}\sim B}[p(\mathbf{b})] = E_{\mathbf{b}\sim B}[SW_{colluders}(\mathbf{b})].
\end{split}
\]
\end{proof}


\begin{lemma}
\label{lem:single_bidder_composition_dsic}
In the single-bidder case, the result of an IC collusion $c,t$ against a DSIC auction $a, p$, is a DSIC auction with $a' = a \circ c, p' = p \circ c - t$. 
\end{lemma}
\begin{proof}
It is immediate that the allocation and payment rules follow the above form. The DSIC property follows from the collusion IC condition when $c$ is not the identity function 
, and from the original auction DSIC condition otherwise. 

\end{proof}


\begin{prop}
The deterministic DSIC auctions for a single bidder are exactly the class of posted prices with arbitrary burn. I.e., for some $v^*\geq 0$, $\tilde{x}(b_1) = (b_1 \geq v^*), \tilde{p}(b_1) = v^*$, and $\tilde{\beta}(b_1) \leq v^*$ (but has no further restrictions). 
\end{prop}
\begin{proof}
The posted-price characterization follows from Myerson's characterization of DSIC auctions as the ones with monotone allocations and payment rules following ``critical bid''. Monotone allocation rules in the single-bidder case are of the form $a(b) = 1[b\geq v^*]$ for some $v^* \in R_+$. Then, this determines that $p(b) = 1[b\geq v^*] \cdot v^*$. This can be implemented as a posted price where the seller offers the item at price $v^*$, and the bidder decides whether to accept. 

The burning rule has no affect on the buyer but only on the seller's revenue and can be arbitrary. 
\end{proof}

When considering IC+IR collusion, even in the restricted single-bidder case, we need to specify the outcomes in the multi-bidder case, as they are baked into the IC collusion condition. Since our focus at this point is on the single bidder case, the straight-forward assumption is that the protocol does not include any transaction in the multi-bidder case. 

To better understand the differences between the IC and IR conditions for collusion notions, we start by considering IC-collusion (that are not necessarily IR). Here, we see the characterization exactly matches the characterization for DSIC+OCA-proof single bidder auctions in \cite{gafni2024barriers}. Thus, adding the IC constraint over possible collusion does not result in a wider class of auctions. Technically, unlike the proof of the corresponding Lemma~3.6 of \cite{gafni2024barriers}, which uses an analogy to Myerson's argument for payments, we give an explicit construction of the collusion. 

\begin{prop}
\label{prop:single_bidder_ic_collusion}
The deterministic DSIC and (IC)-collusion resistant auctions for a single bidder are exactly the class of ``posted burn'' auctions (\cite{gafni2024barriers}). I.e., for some $b^* \geq 0$, $\tilde{x}(b_1) = (b_1 \geq b^*), \tilde{p}(b_1) = b^*$, and $\tilde{\beta}(b_1) = b^*$. 
\end{prop}
\begin{proof}
Our starting point is the space of DSIC single-bidder auctions, as characterized by proposition~\ref{prop:single_bidder_ic_collusion}. Thus, consider an auction of the form $\tilde{x}(b_1) = (b_1 \geq v^*), \tilde{p}(b_1) = v^*$, for some $v^*\geq 0$, with some arbitrary burning rule $\beta$. 

Next, we wish to limit ourselves to the class of posted prices with constant burn. Assume that $|Image(\beta)| > 1$, and there is some $b'_H, b'_L$ so that $\beta(b'_H) = B^H \geq B^L = \beta(b'_L)$. The following collusion increases the joint utility:
$$c(b_1) = \begin{cases} b'_L & b_1 = b'_H \\ b_1 & Otherwise\end{cases}, t(\mathbf{b}) = (0,0).$$


The collusion is incentive-comaptible for bidder $1$. This follows for any $v_1 \neq b'_H$ due to incentive-compatibility of the posted-price mechanism (i.e., there are no transfers and the bidder gets $u_1(v_1 ; v_1)$
, and for $v_1 = b'_H$ it follows since $$u_1(c(b'_1) ; b'_H) = 1[c(b'_1) \geq v^*] \cdot (b'_H - v^*) \leq b'_H - v^* = 1[b'_L \geq v^*] (b'_H - v^*) = 1[c(b'_H) \geq v^*] (b'_H - v^*). $$
The collusion is incentive-compatible for the miner since adding or removing any bids results in utility $0$. 

Finally, we wish to show that the posted price must equal the constant burn. Assume towards contradiction that the auction is of the form 
$v^* > b^* \geq 0$, $\tilde{x}(b_1) = (b_1 \geq v^*), \tilde{p}(b_1) = v^*$, and $\tilde{\beta}(b_1) = b^*$. The following collusion increases the joint utility: $$c(b_1) = \begin{cases} v^* & b_1 \geq b^* \\ b_1 & Otherwise\end{cases}, t(\mathbf{b}) = \begin{cases} (v^* - b^*, b^* - v^*) & b_1 \geq b^* \\ (0,0) & Otherwise\end{cases},$$

and is incentive-compatible for the bidder by Lemma~\ref{lem:single_bidder_composition_dsic}.


The collusion is incentive-compatible for the miner since the utility from the auction when implementing the collusion is non-negative, while adding or removing any bids results in utility $0$. 

Finally, we note that our characterization is exact, as Lemma~3.6 of \cite{gafni2024barriers} shows that the class of auctions we characterize satisfies the more restrictive requirement of DSIC+$1$-OCA. 
\end{proof}

However, when we further require the individually-rational condition for the collusion (i.e., an IC+IR collusion), the last part of the argument in Proposition~\ref{prop:single_bidder_ic_collusion} stops holding generally. We get that the general form of a DSIC+(IC+IR)-collusion is that of a posted-price with constant burn, where the posted-price is not too high as a function of the burn. To provide this characterization, we start by generalizing the Myerson reserve price to incorporate a constant burn specified by the protocol.

\begin{definition}
\textit{Virtual value with constant burn} Let $F(v), f(v)$ be the cumulative and probability density functions of bidder values respectively. Then 
$$\phi_{\beta}(v) = v - \frac{1 - F(v)}{f(v)} - \beta.$$

We say that $F$ is a regular distribution when $\phi_0(v)$ is monotonically increasing in $v$. 

It will be convenient to define, given a posted price $\rho$, $Rev_{\beta}(\rho) = E_{v \sim F}[u_{miner}(v; v)]$, where the auction being considered is the posted price $\rho$ with constant burn $\beta$. 

The Myerson reserve price $\mu(\beta)$ given a burn $\beta$ is:

$\mu(\beta) = \max_{\rho} Rev_{\beta}(\rho) = \max_{\rho} E_{v \sim F}[1[v\geq \rho] \cdot (\rho-\beta)$.
\end{definition}

\begin{lemma}
\label{lem:regular_monotonicity}
For any $\beta$, $\phi_{\beta}(v)$ is monotonically increasing in $v$ if and only if $F$ is regular. 
\end{lemma}

\begin{proof}
This holds since $\phi_0(v)$ is monotonically increasing in $v$ if and only if $F$ is regular by definition, and for any $\beta$, $\phi_0(v) = \phi_{\beta}(v) + \beta$, i.e., the functions differ by a constant. 
\end{proof}

The following proposition immediately follows from applying Myerson's \cite{myerson1981optimal} argument in the presence of a constant burn. 

\begin{prop}
With a constant burn $\beta$, 

$$E_{v \sim F}[u_{miner}(v ; v)] = E_{v \sim F}[\phi_{\beta}(v) \cdot x(v)].$$

When $F$ is a regular distribution, this is maximized exactly when the posted price is set to $\mu(\beta)$, as $\phi_{\beta}(\mu(\beta)) = 0$. When $F$ is not regular, this is maximized when the posted price is set 
\end{prop}

\begin{prop}
$\mu(\beta)$ is monotone non-decreasing with $\beta$. 
\end{prop}
\begin{proof}
For a regular distribution $F$, define $\phi^{-1}(\beta) = \min \{v\}_{\phi_0(v) = \beta}$. Since $\phi_0$ is monotone non-decreasing, its inverse $\phi^{-1}$ is also monotone non-decreasing. But $\phi^{-1}$ is exactly $\mu(\beta)$. 

For a non-regular distribution, notice that the ironing operation averages over different values of the virtual function. Thus, the constant change $\beta$ is constant for all averaged values, and so if $\phi_{\beta}(v) = \phi_0(v) + \beta$, then so is $\tilde{\phi}_{\beta}(v) = \tilde{\phi}_0(v) + \beta$. The above inverse function argument then holds for the ironed virtual function.
\end{proof}

Though the effect of the constant burn $\beta$ on virtual function is a straightforward constant change, the implications for other aspects of optimal pricing are not linear. I.e., it is not true, for example, that given a Myerson reserve price without burn, adding the burn results in the Myerson reserve price with the burn. We illustrate this with the following example:

\begin{example}
Consider the uniform value distribution over the interval $[0,1]$: $V = UNI([0,1])$. We have:
$$\phi_{\beta}(v) = v - \frac{1 - v}{1} - \beta = 2v - 1 - \beta,$$
$$\mu(\beta) = \phi_{\beta}^{-1}(0) = \frac{1 + \beta}{2}$$,

$$Rev_{\beta}(\mu(\beta)) = \int_{\frac{1 + \beta}{2}}^{1} \phi_{\beta}(v) dv = \int_{\frac{1 + \beta}{2}}^{1} (2v - 1 - \beta) dv = (v^2 - (1 + \beta) v) |^1_{\frac{1 + \beta}{2}} = -\beta + \frac{(1 + \beta)^2}{4} = \frac{(1 - \beta)^2}{4} ,$$

while the bidder utility given the posted price $\mu(\beta)$ is:
$$E_{v \sim F}[u_1(v; v)] = \int_{\frac{1+\beta}{2}}^1 (v - \frac{1+\beta}{2})dv = (\frac{v^2}{2} - \frac{1+\beta}{2} v)|^1_{\frac{1+\beta}{2}} = \frac{(1 - \beta)^2}{8}.$$

The following graphs describes the optimal (Myerson) price, together with the expected revenue, bidder utility, and realized burn it yields, for each possible protocol burn rate:

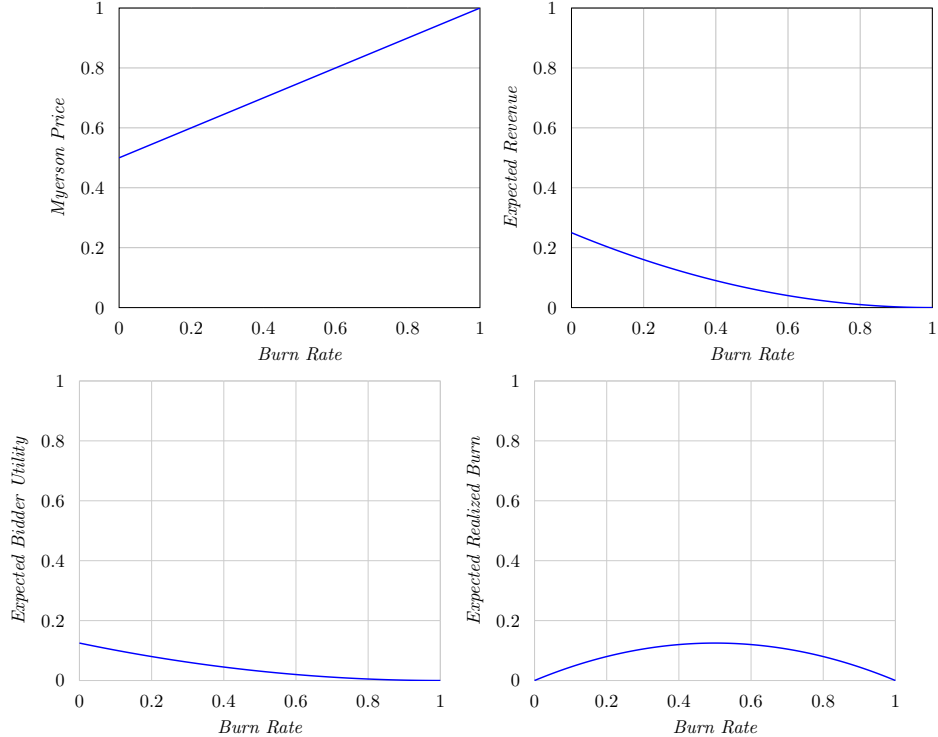
\begin{tikzpicture}[scale=0.7]
    \begin{axis}[
        domain=0:1,
        xmin=0, xmax=1,
        xlabel = Burn Rate,
        ymin=0, ymax=1,
        samples=100,
        ylabel = Myerson Price,
        legend style={
          fill opacity=0.8,
          draw opacity=1,
          text opacity=1,
          at={(1.05,0.5)},
          anchor=west,
          font=\tiny
        },
        tick align=outside,
        x grid style=
        xmajorgrids,
        xmajorticks,
        xtick style={color=white},
        ymajorgrids,
        ymajorticks,
        ytick style={color=white},
        every axis plot/.append style={thick}
    ]

        \addplot+[mark=none] {(1+x)/2};
    \end{axis}
\end{tikzpicture}
\begin{tikzpicture}[scale=0.7]
    \begin{axis}[
        domain=0:1,
        xmin=0, xmax=1,
        xlabel = Burn Rate,
        ymin=0, ymax=1,
        samples=100,
        ylabel = Expected Revenue,
        legend style={
          fill opacity=0.8,
          draw opacity=1,
          text opacity=1,
          at={(1.05,0.5)},
          anchor=west,
          font=\tiny
        },
        tick align=outside,
        xmajorgrids,
        xmajorticks,
        xtick style={color=white},
        ymajorgrids,
        ymajorticks,
        ytick style={color=white},
        every axis plot/.append style={thick}
    ]

        \addplot+[mark=none] {(1-x)*(1-x)/4};
    \end{axis}
\end{tikzpicture}
\begin{tikzpicture}[scale=0.7]
    \begin{axis}[
        domain=0:1,
        xmin=0, xmax=1,
        xlabel = Burn Rate,
        ymin=0, ymax=1,
        samples=100,
        ylabel = Expected Bidder Utility,
        axis line style={lightgray204},
        legend style={
          fill opacity=0.8,
          draw opacity=1,
          text opacity=1,
          at={(1.05,0.5)},
          anchor=west,
          draw=lightgray204,
          font=\tiny
        },
        tick align=outside,
        x grid style={lightgray204},
        xmajorgrids,
        xmajorticks,
        xtick style={color=white},
        y grid style={lightgray204},
        ymajorgrids,
        ymajorticks,
        ytick style={color=white},
        every axis plot/.append style={thick}
    ]

        \addplot+[mark=none] {(1-x)*(1-x)/8};
    \end{axis}
\end{tikzpicture}
\begin{tikzpicture}[scale=0.7]
    \begin{axis}[
        domain=0:1,
        xmin=0, xmax=1,
        xlabel = Burn Rate,
        ymin=0, ymax=1,
        samples=100,
        ylabel = Expected Realized Burn,
        axis line style={lightgray204},
        legend style={
          fill opacity=0.8,
          draw opacity=1,
          text opacity=1,
          at={(1.05,0.5)},
          anchor=west,
          draw=lightgray204,
          font=\tiny
        },
        tick align=outside,
        x grid style={lightgray204},
        xmajorgrids,
        xmajorticks,
        xtick style={color=white},
        y grid style={lightgray204},
        ymajorgrids,
        ymajorticks,
        ytick style={color=white},
        every axis plot/.append style={thick}
    ]

        \addplot+[mark=none] {(1-x)*x/2};
    \end{axis}
\end{tikzpicture}

\end{example}

The Myerson price serves an important role as the maximal price that allows for an (IC+IR)-collusion-resistant posted-price mechanism. When the price goes higher, both bidder and miner benefit from colluding to lower the set price. On the other hand, when the price is lower, at least with regular distributions, the monotonicity of the virtual value guarantees that there is a trade-off between the bidder and the miner's utility, and so they would not collude to change the price. 
This is summarized in our characterization:

\begin{theorem}
\label{thm:single_bidder_regular}
When the bidder value is drawn from a regular distribution, the deterministic DSIC and (IC+IR)-collusion resistant auctions for a single bidder are exactly the class of posted-price $v^*$ mechanisms with a constant burn $b^*$ so that $b^* \leq v^* \leq \mu(b^*)$. I.e., for such $v^*, b^*$ that satisfy this condition, $\tilde{x}(b_1) = (b_1 \geq v^*), \tilde{p}(b_1) = v^*$, and $\tilde{\beta}(b_1) = b^*$. 
\end{theorem}

\begin{proof}
We can start our characterization by examining the class of posted prices with constant burn, as it is straight-forward to follow the proof of Proposition~\ref{prop:single_bidder_ic_collusion} up to that point (by additionally showing that the suggested collusion is individually rational).  

If $v^* > \mu(b^*)$, consider the collusion:
$$c(b_1) = \begin{cases} v^* & b_1 \geq \mu(b^*) \\ b_1 & Otherwise\end{cases}, t(\mathbf{b}) = \begin{cases} (v^* - \mu(b^*), \mu(b^*) - v^*) & b_1 \geq \mu(b^*) \\ (0,0) & Otherwise\end{cases}.$$

We know by Lemma~\ref{lem:single_bidder_composition_dsic} that collusion of this type against a posted-price is incentive-compatible for both the bidder and the miner (as it results in a posted price which is MMIC). Regarding individual rationality, it is IR for the miner as it maps a posted-price mechanism with price $v^*$ and burn $b^*$ to a posted-price mechanism with price $\mu(b^*)$ and burn $b^*$, and by the definition of $\mu(b^*)$ it yields the optimal expected revenue for the miner. It is IR for the bidder as the collusion increases her probability of inclusion (at a non-negative utility), and lowers her payment in all cases she was previously included from $v^*$ to $\mu(b^*)$. 

If $b^* \leq v^* \leq \mu(b^*)$, we know by Proposition~\ref{prop:single_bidder_ic_collusion} that an IC collusion must result in a posted price auction. Since by individual rationality for the bidder, it increases her utility, it must \textit{lower} the posted price. By Proposition~\ref{prop:single_bidder_ic_collusion}, and since the distribution is regular, moving from some $\rho \leq \mu(\beta)$ to a lower posted price means that the miner only allocates more values with negative virtual valuation, i.e., has a lower revenue. Thus the collusion is not IR for the miner. 
\end{proof}

The characterization for non-regular distributions generally follows the same lines, but with the caveat that even prices lower than $\mu(\beta)$ are not necessarily guaranteed to be free of collusion. 

\begin{theorem}
\label{thm:single_bidder_nonregular}
When the bidder value is drawn from a non-regular distribution, the deterministic DSIC and (IC+IR)-collusion resistant auctions for a single bidder are exactly the class of posted-price $v^*$ mechanisms with a constant burn $b^*$ so that $b^* \leq v^* \leq \mu(b^*)$, with the additional constraint that there is no such $v' < v^*$ so that $\phi_{\beta^*}(v') = 0$ and $\int_{v'}^{v^*} \phi_{\beta^*}(v) f(v) dv > 0$. 

I.e., for such $v^*, b^*$ that satisfy this condition, $\tilde{x}(b_1) = (b_1 \geq v^*), \tilde{p}(b_1) = v^*$, and $\tilde{\beta}(b_1) = b^*$. 
\end{theorem}


\begin{example}
\label{ex:nonregular_discontinuity}
Consider a posted price mechanism with zero burn ($\beta^* = 0$). Let $F$ be the non-regular distribution with $Supp F = [0,1.20018]$ such that $F(v) = \frac{5.62}{2}v - \frac{10}{3}v^2 + \frac{5.62}{4}v^3$. We have $f(v) = \frac{5.62}{2} - \frac{20}{3}v + \frac{3 \cdot 5.62}{4}v^2$, and $\phi_0(v) = \frac{1.33333 v^3-2.37248 v^2+1.33333 v-0.237248}{v^2-1.58165v+0.666667}$. 

The monopolistic price that maximizes revenue is $0.845679$, which is the virtual value function's largest root. This can be observed directly by examining the virtual value function graph. It is an implication of Lemma~\ref{lem:regular_monotonicity} is that the monopolistic price is the root $r$ of the virtual value function which maximizes $\int_r^{\infty} \phi(z)f(z) dz$. 

Finally, the collusion-free prices are $[0,0.380043] \cup [0.664978,0.845679]$. For any price $p \in (0.380043, 0.664978)$, the miner and bidder can collude to lower it to $0.380043$, since $\int_{0.380043}^p \phi_0(z) f(z) dz > 0$. 

In Appendix~\ref{app:non_regular} we provide some details on how to explicitly construct such examples. 

\begin{tikzpicture}[scale=0.7]
    \begin{axis}[
        domain=0:1.20018,
        xmin=0, xmax=1.20018,
        xlabel = Value $v$,
        ymin=0, ymax=1,
        samples=100,
        ylabel = Cumulative Density Function $F(v)$,
        axis line style={lightgray204},
        legend style={
          fill opacity=0.8,
          draw opacity=1,
          text opacity=1,
          at={(1.05,0.5)},
          anchor=west,
          draw=lightgray204,
          font=\tiny
        },
        tick align=outside,
        x grid style={lightgray204},
        xmajorgrids,
        xmajorticks,
        xtick style={color=white},
        y grid style={lightgray204},
        ymajorgrids,
        ymajorticks,
        ytick style={color=white},
        every axis plot/.append style={thick}
    ]

        \addplot+[mark=none] {2.81*x - 10/3*x*x + 1.405*x*x*x};
    \end{axis}
\end{tikzpicture}
\begin{tikzpicture}[scale=0.7]
    \begin{axis}[
        domain=0:1.20018,
        xmin=0, xmax=1.20018,
        xlabel = Value $v$,
        ymin=0, ymax=4,
        samples=100,
        ylabel = Probability Density Function $f(v)$,
        axis line style={lightgray204},
        legend style={
          fill opacity=0.8,
          draw opacity=1,
          text opacity=1,
          at={(1.05,0.5)},
          anchor=west,
          draw=lightgray204,
          font=\tiny
        },
        tick align=outside,
        x grid style={lightgray204},
        xmajorgrids,
        xmajorticks,
        xtick style={color=white},
        y grid style={lightgray204},
        ymajorgrids,
        ymajorticks,
        ytick style={color=white},
        every axis plot/.append style={thick}
    ]

        \addplot+[mark=none] {2.81 - 20/3*x + 3*1.405*x*x};
    \end{axis}
\end{tikzpicture}
\begin{tikzpicture}[scale=0.7]
    \begin{axis}[
        domain=0:1.20018,
        xmin=0, xmax=1.20018,
        xlabel = Value $v$,
        ymin=-1, ymax=2,
        samples=100,
        ylabel = Virtual Value Function $\phi_0(v)$,
        axis line style={lightgray204},
        legend style={
          fill opacity=0.8,
          draw opacity=1,
          text opacity=1,
          at={(1.05,0.5)},
          anchor=west,
          draw=lightgray204,
          font=\tiny
        },
        tick align=outside,
        x grid style={lightgray204},
        xmajorgrids,
        xmajorticks,
        xtick style={color=white},
        y grid style={lightgray204},
        ymajorgrids,
        ymajorticks,
        ytick style={color=white},
        every axis plot/.append style={thick}
    ]

        \addplot+[mark=none] {
        (1.33333 * x*x*x - 2.37248*x*x+1.33333*x-0.237248)/(x*x-1.58165*x+0.666667)};      
    \end{axis}
\end{tikzpicture}
\begin{tikzpicture}[scale=0.7]
    \begin{axis}[
        domain=0:1.20018,
        xmin=0, xmax=1.20018,
        xlabel = Value $v$,
        ymin=-1, ymax=2,
        samples=100,
        ylabel = {Adjusted Virtual Value Function $\phi_0(v) f(v)$},
        axis line style={lightgray204},
        legend style={
          fill opacity=0.8,
          draw opacity=1,
          text opacity=1,
          at={(1.05,0.5)},
          anchor=west,
          draw=lightgray204,
          font=\tiny
        },
        tick align=outside,
        x grid style={lightgray204},
        xmajorgrids,
        xmajorticks,
        xtick style={color=white},
        y grid style={lightgray204},
        ymajorgrids,
        ymajorticks,
        ytick style={color=white},
        every axis plot/.append style={thick}
    ]

        \addplot+[mark=none] {
        5.62*x*x*x-10*x*x+5.62*x-1}; 
        \addplot [only marks, mark=o] table {
0.380043 0
0.553638 0
0.845679 0
};

\addplot [only marks, mark=*] table {
0.380043 0
0.845679 0
};

\addplot [only marks, mark=x] table {
0.664978 -0.0322176
};
        \begin{pgfonlayer}{background}
  \fill[color=black!10] (axis cs:0,-1) rectangle (axis cs:0.380043,2);

    \fill[color=black!10] (axis cs:0.664978,-1) rectangle (axis cs:0.845679,2);
\end{pgfonlayer}
    \end{axis}
\end{tikzpicture}
\begin{tikzpicture}[scale=0.7]
    \begin{axis}[
        domain=0.3:0.9,
        xmin=0.3, xmax=0.9,
        xlabel = Value $v$,
        ymin=-0.1, ymax=0.2,
        samples=100,
        ylabel = {Adjusted Virtual Value Function $\phi_0(v) f(v)$},
        axis line style={lightgray204},
        legend style={
          fill opacity=0.8,
          draw opacity=1,
          text opacity=1,
          at={(1.05,0.5)},
          anchor=west,
          draw=lightgray204,
          font=\tiny
        },
        tick align=outside,
        x grid style={lightgray204},
        xmajorgrids,
        xmajorticks,
        xtick style={color=white},
        y grid style={lightgray204},
        ymajorgrids,
        ymajorticks,
        ytick style={color=white},
        every axis plot/.append style={thick}
    ]

        \addplot+[mark=none] {
        5.62*x*x*x-10*x*x+5.62*x-1};
\addplot [only marks, mark=o] table {
0.380043 0
0.553638 0
0.845679 0
};

\addplot [only marks, mark=*] table {
0.380043 0
0.845679 0
};

\addplot [only marks, mark=x] table {
0.664978 -0.0322176
};

\begin{pgfonlayer}{background}
  \fill[color=black!10] (axis cs:0.3,-0.1) rectangle (axis cs:0.380043,0.2);

    \fill[color=black!10] (axis cs:0.664978,-0.1) rectangle (axis cs:0.845679,0.2);
\end{pgfonlayer}
    \end{axis}
\end{tikzpicture}

\end{example}

Our characterization of Theoremes~\ref{thm:single_bidder_regular},~\ref{thm:single_bidder_nonregular} shows that collusion-free prices are bounded from above by the monopolist price. This also means that their \textit{welfare} is bounded from below by the welfare guaranteed by the monopolist price. This allows us to take advantage of results in the algorithmic mechanism design literature \cite{hartlineTextbook}, for example:

\begin{corollary}
Collusion-free prices guarantee at least an $e$-approximation of the optimal welfare with monotone-hazard-rate distributions.
\end{corollary}

However, even with regular distributions (of which monotone-hazard-rate distributions are a subclass), the welfare of collusion-free prices can be arbitrarily bad:

\begin{corollary}
Let $F$ be a regular distribution and let $M = \max Supp F$. Then collusion-free prices can be at least as bad as a $\Omega(\log M)$-approximation of the optimal welfare.
\end{corollary}

The reader may assume this follows from the example of the equal-revenue distribution, if we cap the distribution at $M$ and normalize. However, we comment on why this is not as straightforward, and give the construction in Appendix~\ref{app:non_regular}. 
The equal-revenue distribution has several disadvantages: 
\begin{itemize}
\item Its expected welfare is infinite,
\item All prices guarantee the same revenue, so we might as well tie-break to the side of maximizing welfare. Even if we do not, any choice of finite price would guarantee almost all the welfare. To ``implement'' the revenue-welfare gap we would need an infinite price, i.e., never sell the item.
\end{itemize}

While the technical note of \cite{hartlineTextbook} does indeed point to how these issues could be addressed, it is not clear that any adjustments (or ``small perturbations'') to the equal-revenue distribution would still yield a \textit{regular} distribution. 

Lastly for this discussion, we want to understand the trade-off between welfare and revenue, within the collusion-free prices. At the monopolistic price, revenue is maximized, but welfare can be bad. If we give the item away, welfare is maximized, but the revenue is $0$. A natural question is whether any collusion-free price guarantees at least a good approximation of at least one of the two. Formally,
\begin{definition}
We say that a price $\rho$ has $C$ Welfare-Revenue-Approximation if $C \cdot Rev(\rho) \geq Rev(\mu(0))$, \textbf{or} $C \cdot E_{v\sim F} [v \cdot 1[v \geq \rho]] \geq E_{v\sim F}[v]$. Let $C_{\rho}$ be the infimum of all such $C$. 

Let $C_F = \max \{C_{\rho}\}_{0 \leq \rho \leq \mu(0)}$. Let $C_{Regular}(M) = \max \{C_F\}_{\text{F is regular, } M = \max Supp F}$, $C_{General}(M) = \max \{C_F\}_{M = \max Supp F}$.
\end{definition}

\begin{theorem}
\label{thm:welfare_revenue_bad}
 $C_{General}(M) \in \Omega(\sqrt{\log(M)})$. 
\end{theorem}
\begin{proof}
We show that $C_{General}(M) \in \Omega(\sqrt{\log(M)})$ using the following distribution:  

Consider $n \geq 5$ points of mass $v_1, \ldots, v_n$ so that $\forall 1 \leq i \leq n-1, v_i = 2^{i-1}, v_n = \sqrt{n}2^{n-2}$, with weights $w_1, \ldots, w_n$ so that $\forall 1 \leq i \leq n-1, w_i = (\frac{1}{2})^i, w_n = (\frac{1}{2})^{n-1}$. 

We have
$$Rev(v_i) = \begin{cases}  1 & 1 \leq i \leq n-1 \\ \frac{\sqrt{n}}{2} & i = n \end{cases},  $$

so the revenue maximizing choice is uniquely $v_n$ (under our condition that $n\geq 5$). We also see directly that every $v_i$ is a collusion-free price, since the miner does not have strictly higher revenue at a lower price. 

Now let us consider setting the price at $v_{n-1}$. In terms of revenue, we get a $\frac{2}{\sqrt{n}}$ fraction relative to setting the price at $v_n$. In terms of welfare, we get:
$$E_{v\sim F}[v \cdot 1[v\geq v_{n-1}]] = v_{n-1}w_{n-1} + v_n w_n = \frac{1}{2} + \frac{\sqrt{n}}{2},$$
while the optimal welfare is:
$$E_{v\sim F}[v] = \sum_{i=1}^n v_i w_i = \frac{n-1}{2} + \frac{\sqrt{n}}{2}.$$

So both in terms of revenue and welfare, the approximation is $\Omega(\sqrt{n})$, and since $M = \sqrt{n}2^{n-2}$, this is $\Omega(\sqrt{\log(M)})$. 

\end{proof}
\begin{remark}
The discrete distribution we use for the lower bound is, in fact, a discrete regular distribution (we elaborate on how this is correctly defined in Appendix~\ref{app:discrete_regular}). It is not immediately clear how to translate a discrete regular distribution to a continuous one while maintaining its properties (in this case, the high approximation ratio). It is straightforward, however, to translate it into a continuous non-regular distribution, by replacing each point-mass with a short $\epsilon$-interval.
\end{remark}




\bibliographystyle{plainurl}
\bibliography{bibliography}

\appendix

\label{app:non_regular}
\section{Constructing a Non-Regular Example for Collusion-free Prices}

In our Example~\ref{ex:nonregular_discontinuity} we wished to provide an explicit (and continuous) construction. It turns out that this is not a straight-forward task. One reason is that many of the known non-regular distributions (such as Beta, Gamma, Weibull distributions and so on, see \cite{morgantiNonRegular}) tend to exhibit undesired behavior: They have $\int_0^{\infty} \phi_0(z) f(z) dz = \infty$, and have several points of discontinuity where $\phi_0(z) \rightarrow \infty$. It is thus ideal if we had a procedure to produce probability distributions $F$ so that their corresponding virtual value function (adjusted to the probability density) $\phi_0(v) \cdot f(v)$ follows a certain polynomial $P(v) = a_0 + a_1\cdot v + \ldots + a_n v^n$ of degree $n$. We can write:
$$v \cdot f(v) - 1 - F(v) = (v - \frac{1 - F(v)}{f(v)}) \cdot f(v)= \phi_0(v) \cdot f(v) = P(v).$$
This is an ordinary differential equation, and a solution of it is:
$$F(v) = 1 + a_0 + \frac{a_1}{2} v + \frac{a_2}{3} v^2 + \ldots + \frac{a_n}{n+1} v^n.$$
We let $a_0 = -1$ so that we can have $F(0) = 0$ (i.e., the support of the distribution begins at $0$). For $F$ to be a valid distribution function, we need to require that it is monotone increasing in $v$:
$$f(v) = \frac{a_1}{2} + \frac{2 \cdot a_2}{3} v + \ldots + \frac{n \cdot a_n}{n+1} v^{n-1} \geq 0,$$

and that its support ends at value $M$ so that $F(M) = 1$. 

In our case, we want the polynomial to have an area above the curve, an area under the curve, and then again an area above the curve. Therefore, it must have at least three real roots, and so it must be of degree at least $3$, and of the form:
$$P(v) = c \cdot v^3 - b \cdot v^2 + a \cdot v - 1,$$
for some positive coefficients $a,b,c$. It can be directly verified that to maintain $f(v) \geq 0$, it needs to hold that $\frac{27}{8} \cdot a \cdot c \geq b^2$. It then becomes a matter of trial and error to find such constants that produce Example~\ref{ex:nonregular_discontinuity}.

\label{app:regular_log_approximation}
\section{A Log Approximation Regular Distribution}

Let $$f_{T,\epsilon}(z) = \begin{cases}\frac{1}{(z+1)^2} & 0 \leq z \leq T \\ \frac{1}{\epsilon(T+1)} & T < z \leq T+\epsilon \\
0 & Otherwise.\end{cases}$$

Then,
$$F_{T,\epsilon}(z) = \begin{cases}1 - \frac{1}{z+1} & 0 \leq z \leq T \\ \frac{T}{T+1} + \frac{z-T}{\epsilon(T+1)} & T < z \leq T+\epsilon \end{cases}$$

It can be immediately verified that this is a monotone increasing function on $[0,T+\epsilon]$ that has $F(0) = 0, F(T+\epsilon) = 1$. Next, we derive sufficient conditions to have the monopolistic price at $T$ (i.e., since we have no burn, $\mu(0) = T$). The revenue as a function of posted price $\rho$ is:

$$Rev(\rho) = \begin{cases}\frac{\rho}{\rho+1} & 0 \leq \rho \leq T \\  \frac{\rho(T+\epsilon - \rho)}{\epsilon(T+1)} & T < \rho \leq T+\epsilon \end{cases}.$$

It is clear that $Rev(\rho)$ is monotone increasing in $\rho$ when $\rho \in [0, T]$. As for the second interval, the expression is maximized when $\rho = \frac{T + \epsilon}{2}$. Thus, if $T \geq \frac{T + \epsilon}{2}$, then the expression is monotonically decreasing in $[T,T+\epsilon]$. It is thus a sufficient condition (for the monopolistic price to be at $T$) to have $T \geq \epsilon$. 

We now notice that the optimal welfare has:
$$E_{v\sim F}[v] = \int_0^T v f(v) dv + \int_T^{T+\epsilon} v f(v) dv, $$
while the welfare of selling at the monopolistic price has:
$$E_{v\sim F}[v\cdot 1[v\geq T]] = \int_T^{T+\epsilon} v f(v) dv.$$

So it suffices to calculate the two expressions $\int_0^T v f(v) dv$, and $\int_T^{T+\epsilon} v f(v) dv$.

We have:
$$\int_0^T v f(v) dv = \int_0^T \frac{v}{(1+v)^2} dv = \int_0^T \left( \frac{1}{1+v} - \frac{1}{(1+v)^2} \right) dv = \log (T + 1) - \frac{T}{T+1}, $$

$$\int_T^{T+\epsilon} v f(v) dv = \int_T^{T+\epsilon} \frac{v}{\epsilon(T+1)} dv = \frac{v^2}{2\epsilon(T+1)} |^{T+\epsilon}_T = \frac{2T + \epsilon}{2T + 2},$$

which yields a $\approx \log T$ gap. 

It remains to show that $F$ is a regular distribution. Consider its 
virtual value function:
$$\phi_0(z) = \begin{cases} -1 & 0 \leq z \leq T \\ 2z - (T + \epsilon) & T < z \leq T+\epsilon \end{cases}.$$

As long as $\epsilon \leq 1$, it is monotonically increasing in the value. 

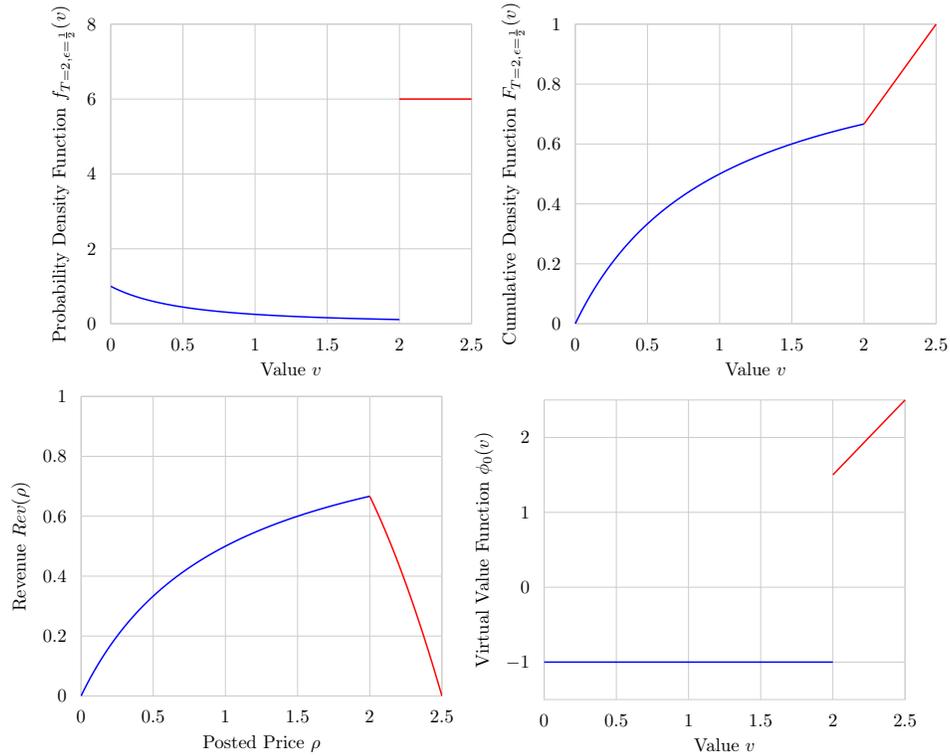
\begin{tikzpicture}[scale=0.7]
    \begin{axis}[
        domain=0:2.5,
        xmin=0, xmax=2.5,
        xlabel = Value $v$,
        ymin=0, ymax=8,
        samples=100,
        ylabel = Probability Density Function $f_{T=2,\epsilon=\frac{1}{2}}(v)$,
        axis line style={lightgray204},
        legend style={
          fill opacity=0.8,
          draw opacity=1,
          text opacity=1,
          at={(1.05,0.5)},
          anchor=west,
          draw=lightgray204,
          font=\tiny
        },
        tick align=outside,
        x grid style={lightgray204},
        xmajorgrids,
        xmajorticks,
        xtick style={color=white},
        y grid style={lightgray204},
        ymajorgrids,
        ymajorticks,
        ytick style={color=white},
        every axis plot/.append style={thick}
    ]

        \addplot+[mark=none][domain=0:2] {1/((1+x)*(1+x))};

        \addplot+[mark=none][domain=2:2.5] {6};
    \end{axis}
\end{tikzpicture}
\begin{tikzpicture}[scale=0.7]
    \begin{axis}[
        domain=0:2.5,
        xmin=0, xmax=2.5,
        xlabel = Value $v$,
        ymin=0, ymax=1,
        samples=100,
        ylabel = Cumulative Density Function $F_{T=2,\epsilon=\frac{1}{2}}(v)$,
        axis line style={lightgray204},
        legend style={
          fill opacity=0.8,
          draw opacity=1,
          text opacity=1,
          at={(1.05,0.5)},
          anchor=west,
          draw=lightgray204,
          font=\tiny
        },
        tick align=outside,
        x grid style={lightgray204},
        xmajorgrids,
        xmajorticks,
        xtick style={color=white},
        y grid style={lightgray204},
        ymajorgrids,
        ymajorticks,
        ytick style={color=white},
        every axis plot/.append style={thick}
    ]

        \addplot+[mark=none][domain=0:2] {1 - 1/(1+x)};

        \addplot+[mark=none][domain=2:2.5] {2/3+2/3*(x-2)};
    \end{axis}
\end{tikzpicture}
\begin{tikzpicture}[scale=0.7]
    \begin{axis}[
        domain=0:2.5,
        xmin=0, xmax=2.5,
        xlabel = Posted Price $\rho$,
        ymin=0, ymax=1,
        samples=100,
        ylabel = Revenue $Rev(\rho)$,
        axis line style={lightgray204},
        legend style={
          fill opacity=0.8,
          draw opacity=1,
          text opacity=1,
          at={(1.05,0.5)},
          anchor=west,
          draw=lightgray204,
          font=\tiny
        },
        tick align=outside,
        x grid style={lightgray204},
        xmajorgrids,
        xmajorticks,
        xtick style={color=white},
        y grid style={lightgray204},
        ymajorgrids,
        ymajorticks,
        ytick style={color=white},
        every axis plot/.append style={thick}
    ]

        \addplot+[mark=none][domain=0:2] {x/(1+x)};

        \addplot+[mark=none][domain=2:2.5] {2/3*x*(2.5-x)};
    \end{axis}
\end{tikzpicture}
\begin{tikzpicture}[scale=0.7]
    \begin{axis}[
        domain=0:2.5,
        xmin=0, xmax=2.5,
        xlabel = Value $v$,
        ymin=-1.5, ymax=2.5,
        samples=100,
        ylabel = Virtual Value Function $\phi_0(v)$,
        axis line style={lightgray204},
        legend style={
          fill opacity=0.8,
          draw opacity=1,
          text opacity=1,
          at={(1.05,0.5)},
          anchor=west,
          draw=lightgray204,
          font=\tiny
        },
        tick align=outside,
        x grid style={lightgray204},
        xmajorgrids,
        xmajorticks,
        xtick style={color=white},
        y grid style={lightgray204},
        ymajorgrids,
        ymajorticks,
        ytick style={color=white},
        every axis plot/.append style={thick}
    ]

        \addplot+[mark=none][domain=0:2] {-1};

        \addplot+[mark=none][domain=2:2.5] {2*x-2.5};
    \end{axis}
\end{tikzpicture}

\label{app:discrete_regular}
\section{Discrete Regular Distributions}
While Myerson \cite{myerson1981optimal} defined regularity for continuous distributions, we could follow the same rationale to define discrete regular distributions. What we would want to maintain is that allocating items to the highest virtual value function (that is above $0$) would maximize revenue \textit{and} remain incentive-compatible. Consider some discrete distribution $F$ with $n$ points of mass $\{v_1, \ldots, v_n\}$ and masses $\{w_1, \ldots, w_n\}$. We can define $s_i = \sum_{i' \geq i} w_i$, and:

$$\phi(v_i) = \begin{cases} v_i - (v_{i+1} - v_i) \cdot \frac{s_{i+1}}{w_i} & 1 \leq i \leq n-1 \\ v_n & i = n\end{cases},$$

and verify that the revenue when setting a posted price at $v_i$ satisfies for any $i$:
\[
\begin{split}
& \sum_{i' \geq i} \phi(v_i) \cdot w_i = \sum_{i \leq i' \leq n-1} \left( v_{i'} w_{i'} - (v_{i'+1} - v_{i'}) s_{i'+1}\right) + v_n \cdot w_n = \\
& \sum_{i \leq i' \leq n-1} \left(v_{i'}s_{i'} - v_{i'+1}s_{i'+1} \right) + v_n s_n = v_i s_i = Rev(v_i).
\end{split}
\]

So indeed the revenue is exactly the sum (adjusted to the weights distribution) of virtual values. A discrete regular distribution is then such that $\phi(v_i)$ is monotonically non-decreasing with $i$. We can directly verify that the distribution in the proof of Theorem~\ref{thm:welfare_revenue_bad} has:
$$\phi(v_i) = \begin{cases} 0 & 1 \leq i \leq n-1 \\ \frac{\sqrt{n}}{2} & i = n\end{cases}.$$

\end{document}